# Homogeneous switching in ultrathin ferroelectric BaTiO$_3$ films

S. Ducharme[1], V. Fridkin[2], R. Gaynutdinov[2], M. Minnekaev[3], A. Tolstikhina[2] and A. Zenkevich[3]

[1] *Dep. of Physics and Astronomy, University of Nebraska-Lincoln, Lincoln, NE, USA*
[2] *Institute of Crystallography Russian Academy of Sciences, Moscow, Russia*
[3] *NRNU "Moscow Engineering Physics Institute", Moscow, Russia*

Switching kinetics in ultrathin epitaxial BaTiO$_3$ films confirms the existence of the homogeneous switching (without domains) in ultrathin ferroelectric films. We suppose that the homogeneous switching in ultrathin films is a common phenomenon for all ferroelectric materials.

PACS numbers: 77.80.Dj, 77.80.Fm.

From the time of domain discovery, the switching kinetics of ferroelectric crystals and thin films has been explained by the domain nucleation and domain dynamics. The domain switching theory of Kolmogorov-Avramy-Ishibashi (KAI) [1–3] successfully explained the switching kinetics of ferroelectric crystals and even thin films with thickness $l \geq 50$ nm. The book of Tagantsev et al. [4] contains a comprehensive theory of domain switching in the ferroelectric crystals and films.

In the simplest case this extrinsic KAI switching has a nucleation-limited mechanism, being an activated process, and the switching time $\tau$ has an exponential dependence on temperature $T$ and reciprocal electric field $E$ or voltage, as follow:

$$\tau = \tau_0 \exp(a(T)/V) \qquad (1)$$

where $\tau_0$ is constant and $a(T)$ typically has a reciprocal dependence on $T$. Notice that there is no threshold, no true coercive field $E_C$ or coercive voltage $V_C$ in this case, because as the switching field is reduced the switching just becomes slower. Therefore, the effective coercive field obtained from polarization-field hysteresis loops is indeed a function of the measurement frequency.

The discovery of two-dimensional ferroelectrics (the Langmuir-Blodgett copolymer films with thickness $l \sim 1$ nm) [5] and ultrathin monodomain perovskite ferroelectric films with $l \sim 3 - 4$ nm [6, 7] and $l \sim 1.2$ nm [8] (three elementary unit cells) raised a question about switching mechanism and its kinetics in the ultrathin ferroelectric films. In the monodomain perovskite films a domain nucleus in [001] direction has the size of a few nm [4]. In the ferroelectric copolymers it should be a little more. Therefore, in ferroelectric films with the thickness in the range 1–4 nm the realization of KAI mechanism is hardly possible. In other words, the switching mechanism of such ultrathin films is homogeneous and has true threshold field, called intrinsic coercive field $E_C^{int}$ or intrinsic voltage $V_C^{int}$, below which the switching is impossible.

Just after discovery of the ultrathin 1 nm ferroelectric films, the kinetics theory of homogeneous switching was developed by means of Landau-Ginzburg-Devonshire (LGD) theory and experimentally verified for ultrathin ferroelectric copolymer films [9 – 12]. The rate of homogeneous switching, obtained from LGD mean field theory for the first order ferroelectrics [13], has a true threshold intrinsic coercive field $E_C^{int}$ and corresponding voltage $V_C^{int}$ and exhibits critical dependence on the applied voltage $V$, as follows [9]:

$$\tau^{-1} = \tau_0^{-1} \exp(V/V_C - 1) \qquad (2)$$

where $\tau$ is the switching time, $\tau_0^{-1}$ is a constant and $V_C^{int}$ is the intrinsic coercive voltage. The eq.(2) drastically differs from the switching rate in domain dynamics (eq.(1)). The eq.(2) shows that homogeneous switching has a true threshold field $E_C^{int}$ or threshold voltage $V_C^{int}$, below which switching is impossible. The constant $\tau_0^{-1}$ decreases smoothly toward zero as the temperature increases toward the transition temperature. We note that switching is only possible at the field above intrinsic coercive field (intrinsic coercive voltage); the switching rate is imaginary when $E < E_C^{int}$ ($V < V_C^{int}$). In [9, 11, 12] we found that the switching dynamics of the ferroelectric copolymer agrees well with eq.(2) for films thinner than 18 nm (homogeneous intrinsic switching), and with eq.(1) for thicker films (extrinsic domain switching).

Recently, polarization switching without domain formations was observed in ultrathin epitaxial PbTiO$_3$ heterostructures with SrRuO$_3$ bottom electrodes coherently strained to SrTiO$_3$(001) single crystal substrates by performing *in situ* synchrotron X-ray scattering measurements [14]. The switching of the ultrathin PbTiO$_3$ films was induced by changing the oxygen chemical potential in equilibrium with the film surface. In this case, the charge of the ions absorbed on the film surface plays the role of the upper electrode. These results show that in PbTiO$_3$ films with thickness $l \sim 2.5 - 10$ nm the switching is homogeneous and the domain nucleation is suppressed. However, the switching kinetics in [14] has not been investigated and the difference between KAI and LGD dynamics is not revealed.

In the present paper, the switching kinetics for the ultrathin heteroepitaxial BaTiO$_3$ films was investigated and nondomain intrinsic switching was confirmed.

BaTiO$_3$/Pt heterostructures were grown on MgO (001) single-crystal substrates by pulsed laser deposition (PLD) in a home-made setup with the base pressure of P ≈ 10$^{-6}$ Pa by using a YAG:Nd laser (λ=1064 nm) operating in the Q-switched regime (τ=15 ns) with the variable output energy E=50÷200 mJ and repetition rate ν=5-50 Hz [15]. Epitaxial Pt underlayer 10 nm in thickness was grown at T=550°C in ultra high vacuum (UHV) on MgO (001) substrates subjected to UHV annealing at T=600°C prior to deposition. 2-15 nm thick strained heteroepitaxial BaTiO$_3$ films were grown on top of Pt in the same vacuum cycle from the sintered stoichiometric BTO target at T=550 °C at the oxygen pressure P≈10$^{-1}$ Pa and further annealed at T=550°C at the oxygen pressure P≈1 Pa, 30 min. The structural characterization of thus grown $BaTiO_3(001)/Pt(001)/MgO(001)$ heterostructures has been reported previously [15]. The heteroepitaxial growth of $BaTiO_3$ on top of Pt is illustrated by the high resolution transmission electron microscopy image of the sample cross-section with fast Fourier transform pictures taken at both MgO/Pt and Pt/$BaTiO_3$ interfaces (Fig.1). We note that due to the lattice mismatch $(a_{BaTiO3} - a_{Pt})/a_{Pt} \approx 1.8\%$ an ultrathin heteroepitaxial BTO layer is under biaxial compressive strain, which ensures that ferroelectric polarization is aligned in the direction perpendicular to the surface [16].

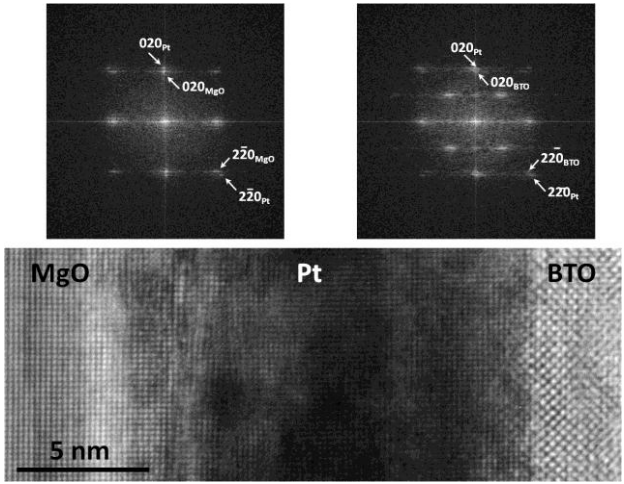

FIG.1. Cross-sectional HRTEM bright-field image of PLD grown BTO/Pt/MgO (100) sample. Fast Fourier transform (FFT) images are taken at MgO/Pt and Pt/BTO interfaces.

The kinetics of switching was investigated by piezoresponse force microscopy (PFM) for 8 nm thick $BaTiO_3$ film. Measurements were carried out with a scanning probe microscope (NTEGRA PRIMA, NT-MDT, Moscow) operating in the contact mode for imaging both topography- by atomic force microscopy (AFM) and relative polarization- by PFM. Sharp tips of soft silicon cantilevers (model CSC21 from MicroMash, Estonia), which were coated with a Ti-Pt conductive layer, have an estimated radius of 40 nm and an estimated imaging resolution of 60 nm. The piezoresponse was imaged by applying the excitation voltage with the amplitude 1.0 V and the frequency 350 kHz to the tip with respect to the bottom Pt electrode, and recording the amplitude and phase of the resulting tip deflection signal from the cantilever position detector with a lock-in amplifier. The piezoresponse image consists of an *x-y* map of the composite PFM signal, which is equal to the PFM amplitude times the sine of the PFM phase. The measurements were carried out in air in a class 100 clean room maintained at temperature $T=24\pm0.05$ °C and relative humidity $40\pm1\%$. The sample temperature was held at $T=24$ °C for all reported measurements.

The hysteresis loops (Fig. 2) were recorded at a fixed location of the film by a pulse-measure method, which consisted of applying a sequence of 1 sec. long voltage pulses of varying amplitude and measuring the piezoresponse at zero voltage. The sequence of pulse amplitude started at −5 V increasing by 0.05 V steps to

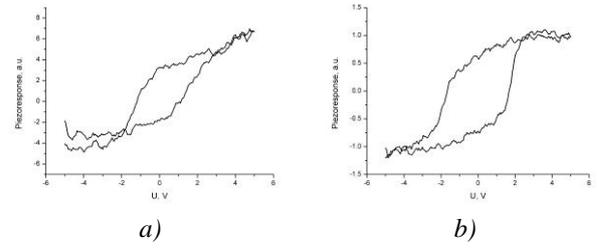

Fig. 2. Hysteresis loops for $BaTiO_3$ measured with PFM: a) Hysteresis loop for $BaTiO_3$ film (8 nm); b) Hysteresis loop for $BaTiO_3$ (001) crystal (1 mm).

+5 V, then back to −5 V. Figure 2a shows the hysteresis loop for the 8 nm thick BTO film.

The coercive voltage which was obtained by measuring the half width of the hysteresis loop, was found $V_C = 1.5\pm0.2$ V. Assuming that the threshold switching field is the field directly underneath the tip, the corresponding coercive field $E_C \approx 3\pm0.2\cdot10^8$ V/m, which satisfactory agrees with the intrinsic coercive field in $BaTiO_3$ $E_C^{int} = P_S/\varepsilon\varepsilon_0 \approx 2\cdot10^8$ V/m [17]. Figure 2b shows hysteresis loop for $BaTiO_3$ single crystal (001) with thickness 0.1 cm. For the single crystal the half width of the hysteresis loop is $V_C=1.7\pm0.2$ V. The corresponding coercive field at $\varepsilon_c \approx 200$ and $\varepsilon_a \approx 6000$ [18] ($\varepsilon_c$, $\varepsilon_a$ are dielectric constants) is $E_C^{ext} \approx 10^6$ V/m [19].

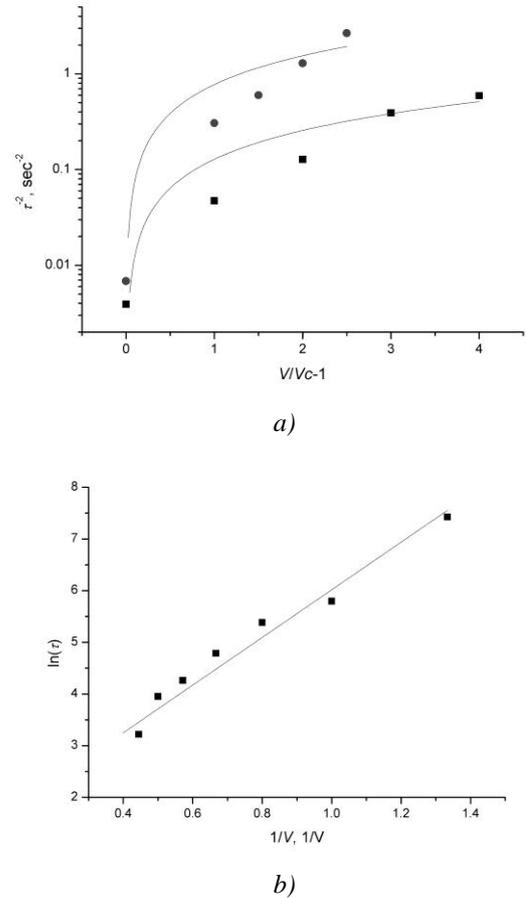

Fig. 3. The kinetics of switching of the 8-nm thick $BaTiO_3$ film (curves 1, 2) (a) and $BaTiO_3$ crystal (b).

The switching kinetics was measured at a fixed point on the film as follows: each measurement was preceded by applying U = –5 V to the tip for 100 seconds to saturate the film in the negative polarization state. Then, a positive pulse of a given voltage and duration was applied to set a new polarization state. Finally, the piezoresponse for the new polarization state was measured at zero bias. The switching time $\tau$ for a given pulse amplitude was defined as the value of the pulse duration for which the resulting piezoresponse crossed zero, and the switching rate was calculated from $1/\tau$.

The distinction between the intrinsic and extrinsic behavior is well illustrated in Fig. 3 which shows the switching rate as a function of the bias voltage for the ultrathin epitaxial $BaTiO_3$ film (a) and $BaTiO_3$ single crystal (b). Figure 3a shows the dependence $\tau^{-2}$ as function of $(V/V_C - 1)$ for the switching in positive (1) and negative (2) directions. The experimental data are satisfactory fitted with the curves from eq.(2), showing that the coercive field $E_C = E_C^{int}$. On the contrary, the switching kinetics for the single crystal (Fig. 3b) agrees with eq.(1) and switching occurs at $V_C^{ext} < V < V_C^{ext}$.

The results of the measurements of the switching kinetics confirm the intrinsic (nondomain) switching in the ultrathin ferroelectric films [9 – 12, 14]. At transition from crystals and thick ferroelectric films to the ultrathin films the KAI nucleation mechanism is replaced by the homogeneous LGD mechanism. We suppose that this transition is a common phenomenon for all ferroelectric materials and the thickness transition criterion is the size of the critical domain nucleus.

The authors are grateful to Dr. O. Uvarov for the TEM analysis.

The work in Moscow was supported by the Russian Ministry of Education and Science under Contract № 11.519.11.3007. Work at the University of Nebraska was supported by the US Department of Energy (DE-FG02-08ER46498).

[1] A. Kolmogorov. Izv. Akad. Nauk, Ser. Math. **3**, 355 (1937).
[2] M. Avrami. J. Chem. Phys. **8**, 212 (1940).
[3] Y. Ishibashi, Y. Takagi. J. Phys. Soc. Jpn. **31**, 506 (1971).
[4] A. Tagantsev, E. Cross and J. Fousek. Domain in Ferroic Crystals and Thin Films (Springer, 2009).
[5] A. Bune, V. Fridkin, S. Ducharme, L. Blinov, S. Palto, A. Sorokin, S. Yudin and A. Zlatkin. Nature **391**, 874 (1998).
[6] T. Tybell, C. Ahn and J.-M. Triscone. Appl. Phys. Lett. **75**, 856 (1999).
[7] C. Lichtensteiger, J.-M. Triscone, J. Junquera and Ph. Ghosez. Phys. Rev. Lett. **94**, 047603 (2005).
[8] D. Fong, A. Kolpak, J. Eastman, S. Streiffer, P. Fuoss, G. Stephenson, C. Thompson, D. Kim, K. Choi, C. Eom, I. Grinberg and M. Rappe. Phys. Rev. Lett. **96**, 127601 (2006).
[9] G. Vizdrik, S. Ducharme, V. Fridkin and S. Yudin. Phys. Rev. B **68**, 094113 (2003).
[10] A. Geivandov, S. Palto, S. Yudin and L. Blinov. JETP **99**, 83 (2004).
[11] R. Gaynutdinov, S. Mitko, S. Yudin, V. Fridkin and S. Ducharme. Appl. Phys. Lett. **99**, 142904 (2011).
[12] R. Gaynutdinov, S. Yudin, S. Ducharme and V. Fridkin. J. Phys: Condens. Matter **24**, 015902 (2012).
[13] V. Ginzburg. Zh. Eksp. Teor. Fiz. **19**, 36 (1949).
[14] M. Highland, T. Fister, M.-I. Richard, D. Fong, P. Fuoss, C. Thompson, J. Eastman, S. Streiffer and G. Stephenson. Phys. Rev. Lett. **105**, 167601 (2010).
[15] A. Zenkevich, M. Minnekaev, Yu. Lebedinskii, K. Bulakh, A. Chouprik, A. Baturin, R. Mantovan, M. Fanciulli, O. Uvarov. Thin Solid Films, doi:10.1016/j.tsf.2011.10.188 in press (2011).
[16] K.J. Choi, M. Biegalski, Y.L. Li, A. Sharan, J. Schubert, R. Uecker, P. Reiche, Y. B. Chen, X. Q. Pan, V. Gopalan, L.-Q. Chen, D. G. Schlom and C. B. Eom, Science **306**, 1005 (2004).
[17] V. Fridkin and S. Ducharme, Phys. of the Solid State **43**, 1320 (2001).
[18] F. Jona and G. Shirane, Ferroelectric Crystals. McMillan, NY, 1962.
[19] G. Rosenman, P. Urenski, A. Agronin, Y. Rosenwaks and M. Molotskii. Appl. Phys. Lett. **82**, 103 (2003).